**Extending ring polymer molecular dynamics rate theory to reactions with non-separable reactants**

Chen Li[1,2], Liang, Zhang[1], Bin Jiang[2,*] and Hua Guo[1,*]

[1]*Department of Chemistry and Chemical Biology, Center for Computational Chemistry, University of New Mexico, Albuquerque, NM 87131, USA*

[2]*State key Laboratory of Precision and Intelligent Chemistry, School of Chemistry and Materials Science, Department of Chemical Physics, University of Science and Technology of China, Hefei, Anhui 230026, China*

*: corresponding authors: *bjiangch@ustc.edu.cn*, *hguo@unm.edu*

**Abstract**

The ring polymer molecular dynamics (RPMD) rate theory is an efficient and accurate method for estimating rate coefficients of chemical reactions affected by nuclear quantum effects. The commonly used RPMD treatment of gas-phase bimolecular reactions adopts two dividing surfaces, one at the transition state and another in the reactant asymptote, where partition functions of separated reactants can be readily obtained. With some exceptions, however, this strategy is difficult to implement for processes on surfaces or in liquids, because the reactants are often strongly coupled with the extended medium (surface or solvent) and thus non-separable. Under such circumstances, the RPMD rate theory with a single dividing surface (SDS) is better suited. However, most of its implementations adopted Cartesian forms of the reaction coordinate, which, in many cases, are not ideal for describing complex reactions. Here, we present a SDS-based RPMD implementation, which are able to tackle the aforementioned challenges. This approach is demonstrated in four representative reactions, including the gas phase H + $H_2$ exchange reaction, gas phase $CH_3NC$ isomerization, H recombinative desorption from Pt(111), and NO desorption from Pd(111). This implementation, which is applicable to both uni- and bi-molecular reactions, offers a unified treatment of gas-phase and surface reaction rate calculations on the same footing.

## I. Introduction

Accurate rate coefficients for elementary reactions play a central role in understanding reaction networks in many important settings, including combustion,[1] atmospheric chemistry,[2] and heterogeneous catalysis.[3] While kinetics of gas phase elementary reactions are a well-established field in both theory and experiment, reaction rates for surface processes are still very difficult to measure and compute accurately. Recently, significant progress has been made in measuring surface reaction rates using velocity imaging.[4] Interestingly, some of the experiments revealed significant nuclear quantum effects even at high temperatures.[5, 6] thus posing a challenge to theoreticians to develop quantitatively accurate rate theory. Such predictive theories should be able to not only describe classical systems, but also be capable of handling quantum effects, such as tunneling and zero-point energy.

The transition-state theory (TST)[7, 8] is arguably the most popular approach for predicting reaction rates, thanks to its simplicity and intuitive physical picture. TST assumes that the reactants and transition states are at thermal equilibrium and the thermal flux passing through the dividing surface completely determines the rate coefficient. Neither recrossing of the dividing surface nor quantum tunneling is included in the traditional TST framework, making it inherently approximate. Attempts to include these effects have been made,[9] but a systematic and reliable solution has still not be achieved.

The formally exact solution to rate calculations is given by the quantum reactive flux theory (RFT) in terms of the flux-side (or another related) correlation function.[10, 11] Various approximations to RFT have been developed, ranging from path-integral based approaches like centroid molecular dynamics (CMD)[12-16] and ring-polymer molecular dynamics (RPMD),[17-21] to the classical Wigner model[22] and purely statistical techniques such as the quantum instanton (QI) model.[23-28] These extensions strive to include nuclear quantum effects with varying levels of sophistication.

In this work, we focus on the RPMD rate theory, which is a path-integral based method to approximately calculate the Kubo-transformed quantum correlation function.[17] With the ring-polymer Hamiltonian,[10, 11] Manolopoulos and coworkers showed that quantum mechanical rate coefficients can be obtained approximately using Newtonian mechanics.[18-21] This approach has been successfully applied to many gas-phase reactions and validated by comparing with quantum benchmarks.[29] The RPMD rate theory comes in two flavors based on the number of dividing surfaces utilized. For gas-phase bimolecular reactions, the isolated reactants in the asymptotic limit permit an analytical expression of the partition function. As a result, the rate coefficient can be readily written in terms of a reaction coordinate (RC) defined by two dividing surfaces, placed in the reactant asymptote and at the transition state, respectively.[21] The short-time rate coefficient can then be straightforwardly determined by the free energy barrier along the RC, which is then corrected by the transmission coefficient accounting for recrossing. Very recently, this double dividing surface (DDS) approach has been extended by some of the current authors to study $H_2$ dissociation/recombination on metal surfaces, achieving results that agree well with experiment.[6, 30, 31]

The RPMD rate theory can also be formulated with a single dividing surface (SDS), located at the transition state.[20] Directly based on the TST framework, this formulation is intuitive and has some advantages, since the reactants are not required to be isolated, as in the DDS implementation. Furthermore, it is capable of handling multichannel reactions by directly introducing multiple RCs.[32] For these reasons, the SDS approach is particularly suitable for processes where the reactants are strongly interacting with their surroundings, thus not separable. A handful of rate calculations have already been reported using the SDS form, focusing on surface reactions[33-36] and condensed phase electron and proton transfer processes mainly involving single species.[20, 37, 38] Thus far, most of these SDS RPMD applications define the RC as either a Cartesian coordinate or a bond length in order to simply the calculation of the velocity through the dividing surface.[20, 33-36] For many complex reactions, however, the RC may not be easily expressed in such simple geometric terms.[20, 39, 40]

In this work, we present a simple and accurate SDS RPMD approach to provide a unified characterization of elementary reactions in both the gas-phase and on surfaces. The key is to numerically compute the thermal flux for any RC appropriate to the reaction. Notably, this approach is applicable to reactions with both sparable and non-separable reactants and does not impose any constraints on the form of the RC. We demonstrate its applicability and accuracy for several representative first- and second-order gas-phase and surface reactions, including the $H+H_2$ exchange reaction,[41] $CH_3CN$ isomerization,[42, 43] recombinative desorption of the adsorbed hydrogen atoms from Pt(111),[5, 31] and NO desorption from Pd(111).[35, 44] The thermal rate coefficients for these different types of reactions are all accurately predicted. We anticipate that this general method is applicable to more complex systems.

## II. Ring Polymer Molecular Dynamics Rate Theory

The starting point of the RPMD rate theory is RFT of Miller and coworkers,[10, 11] in which the rate coefficient is defined as,[45]

$$k(T) = \frac{c_{fs}(t \to \infty)}{Q_r} \tag{1}$$

where $c_{fs}(t \to \infty)$ is the flux-side correlation function and $Q_r$ is the reactant partition function. For gas phase bimolecular reactions, the rate coefficient can be readily expressed in the DDS implementation in terms of two dividing surfaces, one of which is placed in the reactant asymptote where the interaction is zero.[21, 46] The DDS treatment is ideally suited for reactions with separable reactants, but encounters difficulties when the reactants are not separable.

A solution to the problem associated with the DDS implementation of the RPMD rate theory is to avoid the dividing surface in the reactant asymptote. In this SDS formulation,[20] the sole dividing surface $s(\mathbf{x})$ is defined by the RC, $\xi(\mathbf{x})$,

$$s(\mathbf{x}) = \xi(\mathbf{x}) - \xi^{\ddagger}, \tag{2}$$

where, $\xi^{\ddagger}$ is the reference RC, often located at the transition state. The avoidance of the asymptotic dividing surface allows the SDS approach to handle both separable and non-separable reactants, thus offering a unified treatment of reactions in both the gas phase and in condensed phases. So far, the SDS approach has not been tested for gas phase uni- and bi-molecular reactions.

In practice, the rate coefficient is decomposed based on the Bennett-Chandler factorization[47, 48] into

$$k(T) = \kappa(t \to \infty) k_{\text{TST}}(T). \tag{3}$$

Here, $k_{\text{TST}}(T) = c_{fs}(t \to 0^+)/Q_r$ is the so-called transition-state theory (TST) rate coefficient in the short-time ($t \to 0^+$) limit and $\kappa(t \to \infty) = c_{fs}(t \to \infty)/c_{fs}(t \to 0^+)$ is the long-time ($t \to \infty$) limit of the time-dependent transmission coefficient, which accounts for the dynamical recrossing effect. The latter can be numerically calculated by sampling the constrained ensemble at the dividing surface to estimate the ratio of the statistical averages of the initial time derivative or velocity of the dividing surface, $v_s(\mathbf{p},\mathbf{x}) = ds(\mathbf{x})/dt$, for the trajectories reaching the product side in the long- and short-time limits, following

$$\kappa(t \to \infty) = \frac{c_{fs}(t \to \infty)}{c_{fs}(t \to 0^+)} = \frac{\langle \delta[s(\mathbf{x})] v_s(\mathbf{p},\mathbf{x}) h[s(\mathbf{x}_t)] \rangle}{\langle \delta[s(\mathbf{x})] v_s(\mathbf{p},\mathbf{x}) h[v_s(\mathbf{p},\mathbf{x})] \rangle}, \tag{4}$$

where $\delta$ is the Dirac delta function to constrain the ensemble at the dividing surface and $h$ is the Heaviside function to project trajectories onto the product side of the dividing surface. The notation $\langle \ldots \rangle$ stands for an average over the canonical ensemble.

In some previous SDS implementations,[20, 33-36] $v_s(\mathbf{p},\mathbf{x})$ is independent of the configuration $\mathbf{x}$ by requiring the RC to be Cartesian or distance coordinates, thereby allowing $k_{\text{TST}}(T)$ to be expressed in terms of the potential of mean of force (PMF) along the RC. As mentioned in Introduction, this practice can sometime be restrictive when the RC is better represented by non-Cartesian coordinates.

To avoid this limitation, $k_{\text{TST}}(T)$ can be recast with the factor $\langle \delta[s(\mathbf{x})]\rangle$ multiplied in both numerator and denominator:

$$k_{\text{TST}}(T) = \frac{c_{fs}(t \to 0^+)}{Q_r} = \frac{\langle \delta[s(\mathbf{x})]v_s(\mathbf{p},\mathbf{x})h[v_s(\mathbf{p},\mathbf{x})]\rangle}{\langle \delta[s(\mathbf{x})]\rangle} \frac{\langle \delta[s(\mathbf{x})]\rangle}{\langle h[-s(\mathbf{x})]\rangle/V} \quad (5)$$

A benefit of this scheme is that the first term on the right-hand side can be numerically calculated by sampling the constrained ensemble at the dividing surface to statistically average the initial velocity $v_s(\mathbf{p},\mathbf{x}) = \frac{d\xi(\mathbf{x})}{d\mathbf{x}}\frac{d\mathbf{x}}{dt}$ of the trajectories toward the product side at the short-time limit, regardless of the form of the RC. The configuration dependency of $v_s(\mathbf{p},\mathbf{x})$ required by the RC is accurately obtained by numerically sampling. Practically, it can be directly extracted from the transmission coefficient calculation mentioned above. This scheme eliminates the restriction on the RC form present in some previous SDS implementations,[20, 33-36] without increasing the computational cost. At the same time, based on the statistical relationship between the probability density that the RC take the value $\xi'$ and PMF along the RC $F(\xi)$, $\langle \xi(\mathbf{x})-\xi'\rangle = \exp[-\beta F(\xi')]$,[48] the second term can also be further expressed in terms of the PMF, giving

$$k_{\text{TST}}(T) = V \frac{\langle \delta[s(\mathbf{x})]v_s(\mathbf{p},\mathbf{x})h[v_s(\mathbf{p},\mathbf{x})]\rangle}{\langle \delta[s(\mathbf{x})]\rangle} \frac{e^{-\beta F(\xi^\ddagger)}}{\int_{\xi^{(0)}}^{\xi^\ddagger} e^{-\beta F(\xi)}d\xi} \quad (6)$$

The reactant partition function can be fully taken into account by integrating the PMF from the reactant side ($\xi^{(0)}$) along the RC up to the dividing surface ($\xi^\ddagger$). The PMF can be obtained via a standard enhanced sampling technique, as discussed below.

The reactant zone volume ($V$) in Eq. (5) is determined by the type of reaction being studied, which gives rise to the unit of the corresponding rate coefficient.[45] For a first-order reaction, $V$ is simply unity, leading to a unit of s$^{-1}$ for the rate coefficient. For a second-order reaction in the gas phase, $V$ is a spherical shell, $\frac{4\pi}{3}\left[\left(\xi^{\ddagger}\right)^3 - \left(\xi^{(0)}\right)^3\right]$, and the corresponding rate coefficient has a unit of cm$^3$s$^{-1}$. For a second-order reaction on a surface, on the other hand, $V$ becomes a ring on the surface $\pi\left[\left(\xi^{\ddagger}\right)^2 - \left(\xi^{(0)}\right)^2\right]$ and $k$ is given in cm$^2$s$^{-1}$.

Finally, the RPMD version of the RFT can be readily obtained by replacing the classical Hamiltonian with its ring-polymer counterpart, in which the momentum and position vectors of the atoms in the system are replaced with those of the centroids of the corresponding ring polymers.[19] This is based on the well-known isomorphism between the quantum statistical properties and that of a harmonically connected necklace, namely the ring polymer.[49] It has been shown that the RPMD rate theory is capable of approximately accounting for nuclear quantum effects, including zero-point energy and tunneling.[50] When only one bead is used, the RPMD Hamiltonian is reduced to a Newtonian one, and the rate coefficient corresponds to the classical counterpart.

We note in passing that the RPMD rate coefficient is independent of the choice of the dividing surface, as long as it separates the reactant from the product. This is because the static and dynamic factors in Eq. (3) counterbalance each other.[20] This desirable attribute represents an important distinction from TST.

### III. Computational Details

To verify the accuracy of this SDS RPMD method, we selected four representative first- and second-order elementary reactions in the gas phase and on surfaces. These reactions include the H + H$_2$ exchange reaction,[41] CH$_3$NC isomerization,[42, 43] H recombinative desorption from Pt(111),[5, 31] and NO desorption from Pd(111).[35, 44] All calculations were performed using a heavily modified version of the RPMDrate program.[41] A time step of 0.1 fs was selected in all propagation. The required PMFs were computed by umbrella sampling[51] or umbrella integration,[52] in windows biased by harmonic potentials with a force constant $k_i$. The sampling of the constrained ensemble at the dividing surface was implemented with the RATTLE algorithm.[53] The thermal equilibration process was simulated using the Andersen thermostat.[54] The other relevant parameters are summarized in Table 1. Among them, the RC definition is given for each system. The units of all listed time parameters in Table 1 are ps. The listed bead numbers were tested for convergence.

**Table 1: Parameters used in RPMD rate calculations for the four reactions (* denotes the adsorbed state for a species on a surface).**

| Parameters | Reactions | | | |
|---|---|---|---|---|
| | H+H$_2$→H$_2$+H | CH$_3$NC→CH$_3$CN | NO$^*$→NO | H$^*$+H$^*$→H$_2$ |
| $T$ (K) | 300 - 1000 | 472.55 - 532.95 | 600 - 800 | 653 - 953 |
| Beads | 1, 32 | 1, 8 | 1 | 1, 16 |
| RC | $r_2 - r_1$ | $\theta_{CNC}$ | $h_{NO}$ | $r_{HH} - h_{HH}$ |

|  | (bohr) | (rad) | (bohr) | (bohr) |
| --- | --- | --- | --- | --- |
| Umbrella sampling parameters | | | | |
| RC range | (-20,1) | (0,$\pi$) | (1,14) | (-4,4) |
| Window interval | 0.2 | 0.05 | 0.2 | 0.05 |
| $k_i$ | $2\times10^{-4}T$ (hartree bohr$^{-2}$) | $1\times10^{-2}T$ (hartree rad$^{-2}$) | $2\times10^{-4}T$ (hartree bohr$^{-2}$) | $1\times10^{-3}T$ (hartree bohr$^{-2}$) |
| $t_{equilibration}$ | 10 | 10 | 20 | 20 |
| $t_{sampling}$ | 50 | 50 | 125 | 100 |
| $N_{trajectory}$ | 100 | 50 | 400 | 100 |
| Transmission coefficient calculation | | | | |
| $t_{equilibration}$ | 10 | 10 | 10 | 20 |
| $N_{totalchild}$ | 10000 | 10000 | 128000 | 100000 |
| $t_{childsampling}$ | 2 | 2 | 2 | 2 |
| $N_{child}$ | 10 | 10 | 128 | 100 |
| $t_{child}$ | 0.05 | 1.5 | 10 | 0.15 |

## IV. Results and Discussion

### A. H + H$_2$ exchange reaction

We first test this SDS RPMD method in the simplest bimolecular reaction in the gas phase, namely the H + H$_2$ exchange reaction, by comparing with the DDS RPMD results, which have been extensively validated.[21, 29, 41] To this end, we adopted the Boothroyd–Keogh–Martin–Peterson (BKMP2) potential energy surface (PES).[55] The exchange reaction proceeds over a potential barrier of 0.417 eV. In the SDS approach, the RC is defined as the difference between the lengths of the breaking ($r_2$) and forming ($r_1$) bonds.

Results from the SDS and DDS calculations, including the classical (1 bead) and converged RPMD (32 bead) PMFs and transmission coefficients at 300 K, are compared in Fig. 1(a) and (b). The agreement is excellent for results obtained using either 1 or 32 beads. The much lower RPMD free-energy barrier than its classical counterpart suggests strong tunneling in this reaction. Furthermore, Fig. 2 compares the rate coefficients predicted by these two methods at 300, 600 and 1000 K. The rate coefficients follow closely the Arrhenius limit, thanks to its activated nature of the reaction. Again, the significantly higher RPMD rate coefficients at lower temperatures are a clear indication of the nuclear quantum effect. It is clear from the figure that the SDS approach yields essentially the same results as the DDS approach, suggesting that the two methods are practically equivalent in predicting gas-phase bimolecular reaction rates.

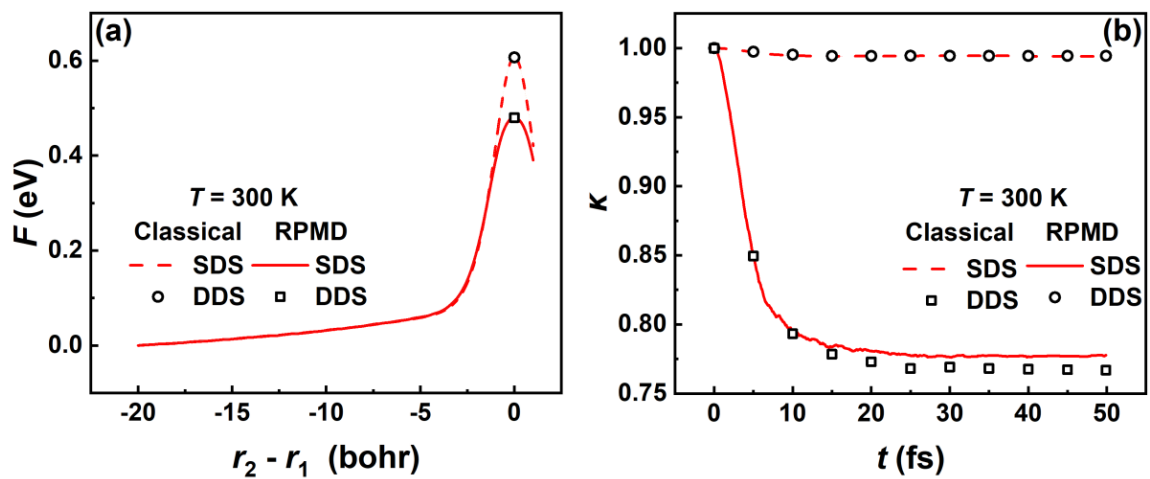

Fig. 1. Comparison of PMFs and transmission coefficients predicted by the SDS and DDS methods with 1 bead (Classical) and 32 beads (RPMD) for the H+$H_2$ reaction at 300 K.

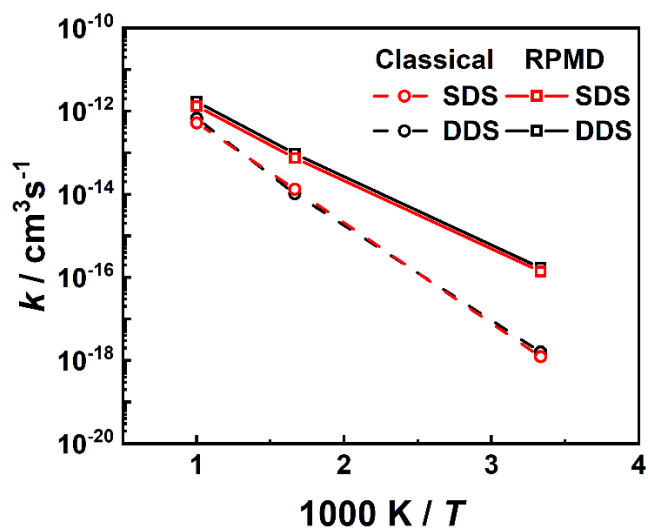

Fig. 2. Comparison of rate coefficients predicted by the SDS and DDS methods with 1 bead (Classical) and 32 beads (RPMD) for the H+H$_2$ reaction at 300, 600 and 1000 K.

## B. CH$_3$NC isomerization

The DDS RPMD rate approach is not amenable to unimolecular reactions because the reactant is not an infinitely separated pair of molecules. Here, a prototypical unimolecular isomerization reaction in the high pressure limit, $CH_3NC \rightarrow CH_3CN$,[42, 43] is chosen as the second gas-phase example for testing the SDS RPMD approach. We adopted the full-dimensional PES recently constructed by Li et al.[43] at the CCSD(T) level of theory, using the permutation invariant polynomial-neural network (PIP-NN) approach.[56, 57] This PES features two wells corresponding to the CH$_3$NC and CH$_3$CN isomers, which are separated by a barrier of 1.727 eV from the CH$_3$NC minimum.

The selection of an optimal RC in calculating rate coefficients is an important topic and many recent studies have focused on this issue.[58, 59] Since the structure transformation along the isomerization path is mainly associated with the angle $\theta_{CNC}$, it is chosen here as the RC which can be easily done in the present SDS approach. In the recent RPMD calculation of the free-energy profile of the isomerization, the differences of the breaking and forming bond lengths at the reactant and the transition state were used to define two dividing surfaces which in turn define the RC as in the Caracal program.[43, 60] Note that these calculated free-energy barriers are subsequently feed to the traditional TST to roughly estimate rate coefficients without following a rigorous RPMD rate formula.[43, 60] In the discussion below, the angle and bond length based RCs are denoted as RC1 and RC2, respectively. Note that RC1 is not a simple function of Cartesian coordinates.

Fig. 3(a) and (b) show the classical (1 bead) and RPMD (8 beads) PMFs and transmission coefficients for the $CH_3NC$ isomerization process at 472.55 K, respectively. It can be seen that the PMF barrier height drops by only about 30 meV as the number of beads increases from one to eight, indicating a small nuclear quantum effect in the isomerization process. The transmission coefficients are less affected by the nuclear quantum effects and converge to about 0.7 after ~1 ps.

Next, in Fig. 4, the classical and RPMD rate coefficients predicted using both RC1 and RC2 by the current SDS approach are compared with the measured rate coefficients at the experimental temperatures (472.55, 503.55 and 532.95 K).[42] Due to nuclear quantum effects, the classical rate coefficients are about half of the RPMD values, regardless of the RC used in the simulations.

Interestingly, the RPMD results obtained using RC1 are significantly smaller than those obtained using RC2 and align more closely with the experimental data. This suggests that the angle-based RC is better for describing this reaction. To validate this point, a committor analysis[61] was performed for these two RCs. To this end, an equilibrium ensemble (1000) of configurations was first sampled by constraining the system at the free-energy barrier at 472.55 K, and then each of these configurations is assigned 100 momentums from the Boltzmann distribution to evolve classically to generate the commitment probability distribution. As shown in Fig. 5, the resulting distribution for RC1 is peaked closer to 0.5 than that of RC2, clearly indicating that $\theta_{CNC}$ is indeed more suitable for characterizing this isomerization process. Overall, these comparisons demonstrate the accuracy of the present SDS RPMD approach in describing unimolecular reactions and the advantage of not restricting the RC form.

In the same figure, the results of Li et al.[43] are also included. The excellent agreement between their results and the experiment may be a coincidence due to the following two reasons. Firstly, their calculation method cannot be regarded as a rigorous RPMD rate treatment, but rather a modified TST with the calculated PMF barriers and transmission coefficients calculated at the transition-state dividing surface. Furthermore, they multiplied the rate coefficient by a factor of 3 due to the $C_{3v}$ symmetry of the $CH_3NC$ molecule, which is incorrect because the equivalent configurations have already been taken into account when performing the umbrella sampling to obtain the free energy. When this factor was removed, the agreement with experiment deteriorates. Consequently, a meaningful comparison with our results is difficult.

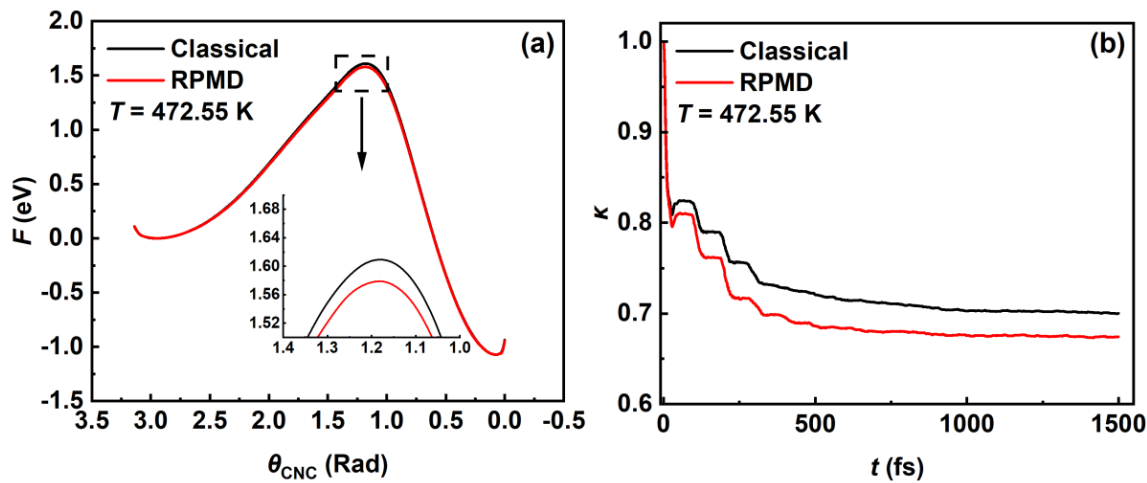

Fig. 3. PMF profiles (a) and transmission coefficients (b) predicted by the SDS method for CH$_3$NC isomerization reaction with 1 bead (Classical) and 8 beads (RPMD) at 472.55 K.

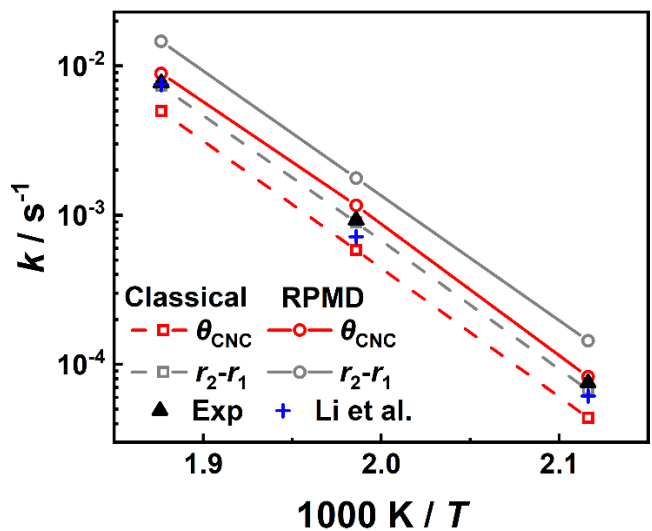

Fig. 4. Comparison between rate coefficients predicted by the SDS method using RC1 ($\theta_{CNC}$) and RC2 ($r_2 - r_1$) for CH$_3$NC isomerization reaction at 472.55, 503.55 and 532.95 K, where 1 bead (Classical) and 8 beads (RPMD) are adopted, respectively. The experimental results[42] ant calculation results of Li et al.[43] are included for comparison.

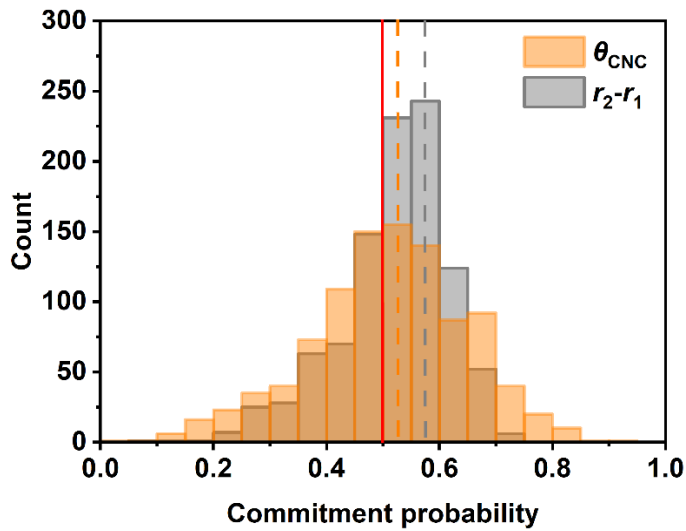

Fig. 5. Committor distribution analysis for reaction coordinates RC1 ($\theta_{CNC}$) (orange) and RC2 ($r_2 - r_1$) (gray) to describe the CH$_3$NC isomerization process at 472.55 K. The red line represents the ideal commitment probability. The orange and gray dash lines correspond to the peaks of the commitment distributions for RC1 and RC2.

### C. Hydrogen recombinative desorption from Pt(111)

The third testing case is provided by the hydrogen recombinative desorption from Pt(111) in the zero-coverage limit, which was the subject of a recent experiment study.[5] This prototypical bimolecular surface reaction has recently been investigated using a DDS version of the RPMD rate theory, assuming the adsorbed hydrogen atoms as a two-dimensional (2D) ideal gas on a rigid Pt(111) surface.[31] This assumption, which is reasonable given the small diffusion barrier of atomic hydrogen on Pt surfaces, allows an analytical expression of the partition function in the reactant asymptote.

Here, we use the SDS implementation of the RPMD rate to eliminate the 2D ideal gas assumption used in our previous work.[31] To this end, we adopted the same PES, which was constructed by the PIP-NN method[56, 62] with DFT points and then adjusted using available experimental data.[31] The minimum energy path for recombinative desorption of two H atoms adsorbed at two adjacent fcc sites features a barrier of 0.73 eV. The RC is defined in terms of the distance between two adsorbed H atoms ($r_{HH}$) and the height of their center of mass from the Pt(111) surface ($h_{HH}$), namely $r_{HH}$ - $h_{HH}$. The surface is treated as rigid.

Fig. 6(a). and (b) show the classical and RPMD PMFs and transmission coefficients for this reaction at 653 K, obtained using the SDS approach. These results are in good agreement with our recent DDS results,[31] which are also included in this figure. The RPMD free-energy barrier is slightly higher than the classical counterpart, due apparently to the zero-point energies of the reactants rather than tunneling, as discussed in our earlier work.[31] In Fig. 7, the calculated rate coefficients are compared with the available experiment[5] and agreement is as good as the DDS results. Based on these results, it can be concluded that surface bimolecular reactions with quantum effects can also be accurately described by the current SDS RPMD approach. This is important because the 2D ideal gas assumption is not applicable to adsorbed species with high diffusion barriers.

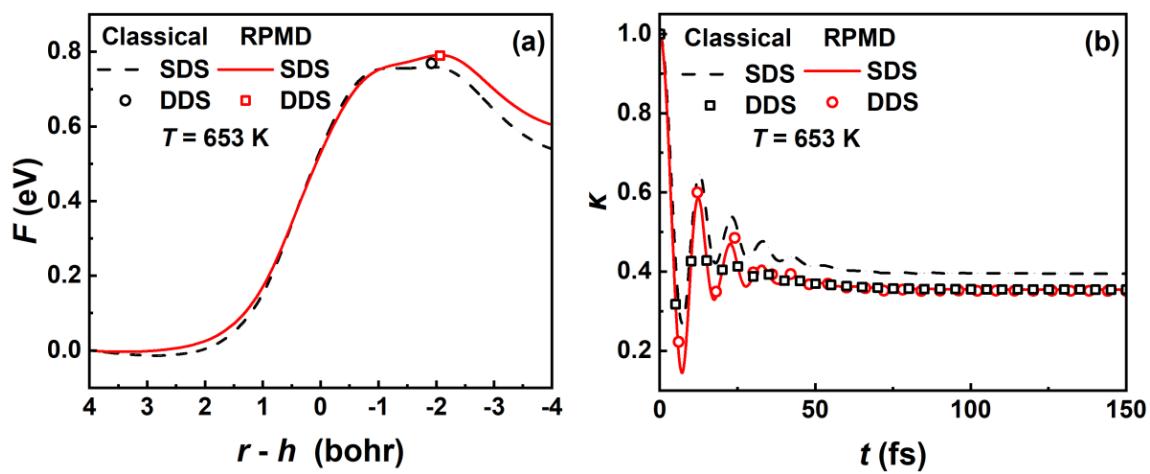

Fig. 6. PMF profiles (a) and transmission coefficients (b) predicted by the SDS and DDS method for the recombination process of hydrogen atoms on Pt(111) with 1 bead (Classical) and 16 beads (RPMD) at 653 K.

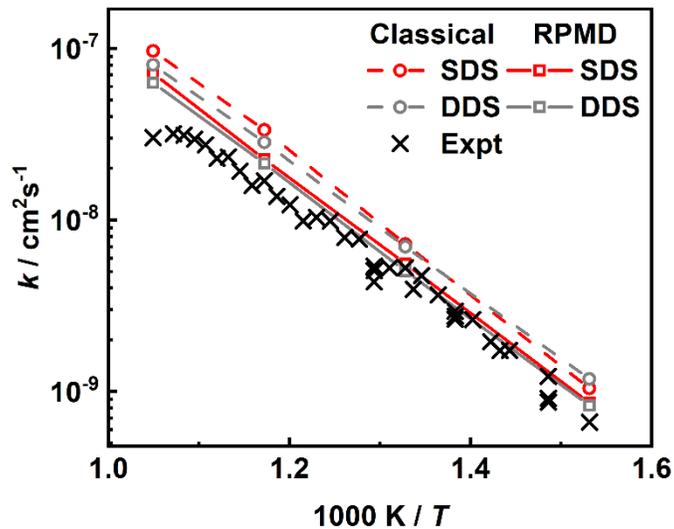

Fig. 7. Comparison between rate coefficients predicted by the SDS method with 1 bead (Classical) and 16 beads (RPMD) and the experimental data[5] for the recombinative

desorption process of hydrogen atoms on Pt(111) at 653, 753, 853 and 953 K. Our previous results based on a DDS approach[31] are also included for comparison.

**D. NO desorption from Pd(111)**

The final example is NO desorption from Pd(111),[35, 44] which has been investigated by some of us using a slightly different SDS approach.[35] The difference lies in the calculation method of the second term in Eq. (6): it was computed numerically in the current work, while in our previous study the velocity at the dividing surface was obtained analytically as the RC was defined in terms of Cartesian coordinates.[35] In contrast to the H recombinative desorption discussed above, the surface atoms are strongly involved in the desorption of NO. The high-dimensional PES employed for this process is the same one developed by us[35] using the embedded atom neural network (EANN) method[63] at the level of the revised Perdew–Burke–Ernzerhof (RPBE) functional. The height of the center of mass of the NO molecule from the Pd(111) surface ($h_{NO}$) was chosen as the RC, as in our previous work.[35] Since NO consists of heavy elements, it is safe to ignore nuclear quantum effects at the experimental temperatures. Consequently, only 1 bead was used in the RPMD calculations. Not surprisingly, exactly the same PMF distribution and transmission coefficients were obtained here and are therefore not shown again.[35] Only final rate coefficients are displayed in Fig. 8. Expectedly, rate coefficients predicted by the present SDS RPMD approach are identical to our previous SDS result, and both are close to the experimental one, within a factor of 3 to 4.[35]

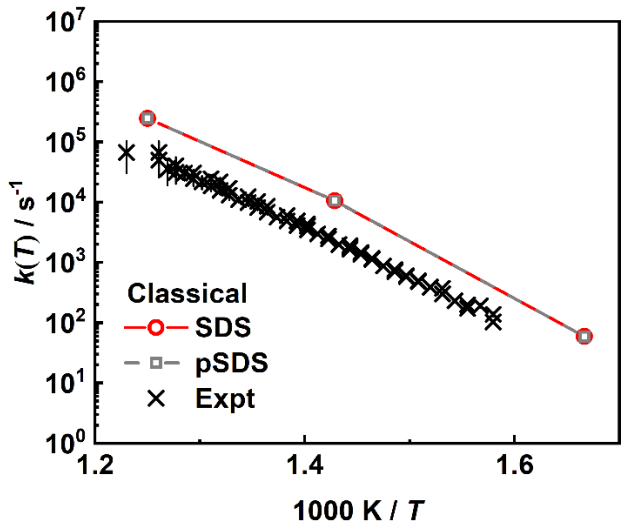

Fig. 8. Comparison between classical SDS rate coefficients and the experimental data as well as our previous SDS results[35] (pSDS) for NO desorption from the Pd(111) surface at 600, 700, and 800 K.

## V. Conclusions

In conventional RPMD rate calculations for activated bi-molecular reactions in the gas phase, two dividing surfaces are typically used by a RC defined in terms of the forming and breaking bond lengths. One dividing surface is placed in the reactant asymptote as it greatly simplifies the calculation. However, such a DDS implementation is not easily amenable to other reactions, particularly those on surfaces or in solutions, because the reactants are often strongly interacting with the medium and thus not separable. Under such circumstances, a SDS implementation is preferred as it uses a single dividing surface at the transition state while the reactant partition function in any cases can be fully considered by the PMF integral. Notably, the SDS approach is applicable to reactions with both separable and non-separable reactants, thus offering a unified treatment. However, many current applications of SDS methods still rely on Cartesian forms of reaction coordinates. In reality, some processes can be better defined using non-Cartesian coordinates. An example is isomerization, in which a bond angle is better suited as the RC.

In this work, we demonstrate that the SDS implementation of RPMD rate theory can seamlessly handle cases involving non-Cartesian RCs, with the conversion factor numerically calculated. This makes it a much more versatile tool for describing a wide range of reaction processes, with either separable or non-separable reactants, on an equal footing. We rigorously tested this methodology on several representative systems, including the H + $H_2$ exchange reaction, $CH_3NC$ isomerization, H recombinative desorption from Pt(111), and NO desorption from Pd(111). For the H + $H_2$ reaction, this method precisely reproduces the PMF barrier, transmission coefficient, and rate coefficient obtained using the well-established DDS implementation of the RPMD rate theory across a broad temperature range. For the $CH_3NC$ isomerization, we demonstrated that an angle-based RC provides a more reliable prediction of the rate coefficients based on committor analysis. In the two surface reactions, the SDS predictions are also found to agree well with experiment and previous theoretical calculations. Importantly, the proper treatment of the reactant state in rate calculations avoids potential errors introduced by approximations such as the 2D ideal gas model.

These compelling results strongly suggest the SDS implementation of the RPMD rate theory is accurate and treats different types of reactions on an equal footing. Given the current interest in nuclear quantum effects in a wide range of chemical processes,[64] it can be expected that this RPMD approach can be directly extended to study more complex reactions in liquids and proteins involving light atoms, further broadening its applicability.

**Conflicts of interest:** There are no conflicts to declare.

**Acknowledgements:** This work at UNM was supported by US National Science Foundation (CHE-2306975 to H.G.). The work at USTC was supported by Strategic Priority Research Program of the Chinese Academy of Sciences (XDB0450101 to B. J.), National Natural Science Foundation of China (22325304 and 22221003 to B. J.). Calculations were performed at the Center for Advanced Research Computing (CARC) at UNM and at the Supercomputing Center of USTC and Hefei Advanced Computing Center. We thank Prof. Jun Li for providing us the $CH_3NC$ isomerization reaction PES and Dr. Yang Liu for thoughtful discussion.